# LOW-DOSE CT DENOISING WITH CONVOLUTIONAL NEURAL NETWORK

*Hu Chen[1, 2], Yi Zhang[1, *], Member, IEEE, Weihua Zhang[1], Peixi Liao[3], Ke Li[1, 2], Jiliu Zhou[1], Senior Member, Ge Wang[4], Fellow, IEEE*

[1]College of Computer Science, Sichuan University, Chengdu 610065, China
[2]National Key Laboratory of Fundamental Science on Synthetic Vision, Sichuan University, Chengdu 610065, China
[3]Department of Scientific Research and Education, The Sixth People's Hospital of Chengdu, Chengdu 610065, China
[4]Department of Biomedical Engineering, Rensselaer Polytechnic Institute, Troy, NY 12180, USA

**ABSTRACT**

To reduce the potential radiation risk, low-dose CT has attracted much attention. However, simply lowering the radiation dose will lead to significant deterioration of the image quality. In this paper, we propose a noise reduction method for low-dose CT via deep neural network without accessing original projection data. A deep convolutional neural network is trained to transform low-dose CT images towards normal-dose CT images, patch by patch. Visual and quantitative evaluation demonstrates a competing performance of the proposed method.

***Index Terms***— Low-dose CT, noise reduction, deep learning, convolutional neural network.

## 1. INTRODUCTION

In the past decades, X-ray computed tomography has been widely used in diagnostic and interventional tasks. With the increasing number of CT scans, the potential radiation risk attracts an increasingly public concern [1]. Most commercial CT scanners utilize the filtered backprojection (FBP) method to analytically reconstruct images. One of the most used methods to reduce the radiation dose is to lower the operating current of the X-ray tube. However, directly lowering the current will significantly degrade the image quality due to the excessive quantum noise caused by an insufficient number of photons in the projection domain.

Many approaches were proposed to improve the quality of low-dose CT images. These approaches can be categorized into three classes: sinogram filtering, iterative reconstruction and image processing.

This work was supported in part by the National Natural Science Foundation of China under Grants 61202160, 61302028 and 61671312, the CSC under Grants 201308510138 and in part by the National Institute of Biomedical Imaging and Bioengineering (NIBIB)/National Institutes of Health (NIH) under Grant R01 EB016977 and U01 EB017140. Asterisks indicate corresponding author.

Sinogram filtering directly smoothens raw data before FBP is applied. Iterative reconstruction solves the problem iteratively, aided by some kinds of prior information on target images. Different priors were proposed, such as in terms of total variation (TV), nonlocal means (NLM) and dictionary learning [2-8]. Despite the successes achieved by these two approaches, they are often restricted in practice due to the difficulty of gaining well-formatted projection data since the vendors are not generally open in this aspect. Meanwhile, the iterative reconstruction methods suffer from heavy computational costs. In contrast to the first two categories, image processing does not rely on projection data, can be directly applied to low-dose CT images, and easily integrated into the current CT workflow. However, it is underlined that the noise in low-dose CT images does not obey a uniform distribution. As a result, it is not easy to remove image noise and artifacts effectively with traditional image denoising methods. Extensive efforts were made to suppress image noise via image processing for low-dose CT. Based on the popular idea of sparse representation, Chen et al adapted K-SVD to deal with low-dose CT images [9]. Also, a block-matching 3D (BM3D) algorithm has been proved powerful in image restoration for different noise types [10].

Recently, deep learning (DL) has generated an excitement in the field of machine learning and computer vision [11]. DL can efficiently learn high-level features from the pixel level through a hierarchical framework. In the field of medical image processing, there are already multiple papers on DL-based image analysis, such as image segmentation, nuclei detection, and organ classification. To our best knowledge, there are few studies proposed for imaging problems. In this regard, Wang et al introduced the DL-based data fidelity into the framework of iterative reconstruction for undersampled MRI reconstruction [12]. Zhang et al. proposed a limited-angle tomography method with deep convolutional neural network (CNN) [13]. Wang shared his opinions on deep learning for image reconstruction [14].

Inspired by the great potential of deep learning in image processing, here we propose a deep convolutional neural network to transform low-dose CT images towards corresponding normal-dose CT images. An offline training stage is needed using a reasonably sized training set. In the second section, the network details are described. In the third section, visual effects and quantitative results are presented. Finally, the conclusion is drawn.

## 2. MEHTHODS

### 2.1. Noise reduction model

Due to the encryption of the raw projection data, post-reconstruction restoration is a reasonable alternative for sinogram-based methods. Once the target image is reconstructed from a low-dose scan, the problem becomes image restoration or image denoising. A difference between low-dose CT image denoising and natural image restoration is that the statistical property of low-dose CT images cannot be easily determined in the image domain. This will significantly compromise the performance of noise-dependent methods, such as median filtering, Gaussian filtering, anisotropic diffusion, etc., which were respectively designed for specific noise types. However, learning-based methods are immune to this problem, because this kind of methods is strongly dependent on training samples, instead of noise type. We model the noise reduction problem for low-dose CT as follows.

Let $\mathbf{X} \in \mathbb{R}^{m \times n}$ is a low-dose CT image and $\mathbf{Y} \in \mathbb{R}^{m \times n}$ is the corresponding normal-dose image, then the relationship can be formulated as:

$$\mathbf{X} = \sigma(\mathbf{Y}), \qquad (1)$$

where $\sigma : \mathbb{R}^{m \times n} \to \mathbb{R}^{m \times n}$ represents the corrupting process due to the quantum noise that contaminates the normal-dose CT image. Thus, the denoising problem can be converted to find a function $f$:

$$f = \arg\min_{f} \| f(\mathbf{X}) - \mathbf{Y} \|_2^2, \qquad (2)$$

where $f$ is treated as the best approximation of $\sigma^{-1}$.

### 2.1. Convolutional neural network

In this study, the low-dose CT denoising problem is solved in the three steps: patch coding, non-linear filtering, and reconstruction. Next, we introduce each step in details.

*2.2.1 Patch encoding*
Sparse representation (SR) is popular for image processing. The key idea of SR is to represent extracted patches of an image with a pre-trained dictionary. Such dictionaries can be categorized into two groups according to how dictionary atoms are constructed. The first group includes analytic dictionaries such as DCT, Wavelet, FFT, etc. The other one is learned dictionaries, which can preserve more application-specific details assuming proper training samples. This SR process can be often implemented as convolution operations with a series of filters, each of which is an atom. Our method is similar in the sense that SR is involved for patch encoding but with a neural network. First, we extract patches from training images with a fixed slide size. Second, the first layer to implement patch coding can be formulated as

$$C_1(\mathbf{y}) = \text{ReLU}(\mathbf{W}_1 * \mathbf{y} + \mathbf{b}_1), \qquad (3)$$

where $\mathbf{W}_1$ and $\mathbf{b}_1$ denote weights and biases respectively, $*$ represents the convolution operator, $\mathbf{y}$ is extracted patch from images, and $\text{ReLU}(x) = \max(0, x)$ is the nonlinear activation function. In CNN, $\mathbf{W}_1$ can be seen as $n_1$ convolution kernels with a fixed size of $s_1 \times s_1$. After patch encoding, we embed the image patches into a feature space, and the output $C_1(\mathbf{y})$ is a feature vector, whose size is $n_1$.

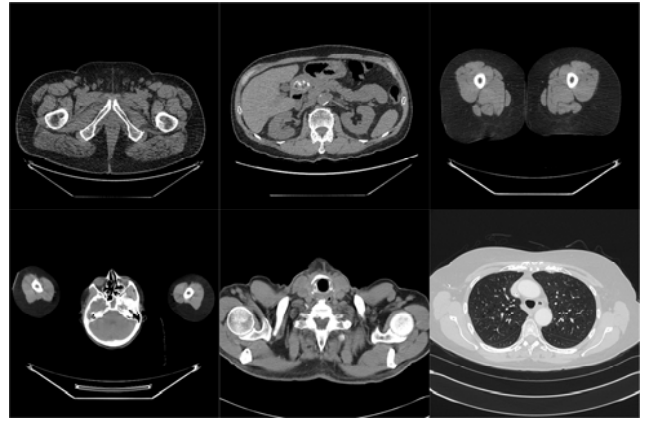

Fig. 1. Typical CT images used for training.

*2.2.2 Non-linear filtering*
After processed by the first layer, a $n_1$-dimensional feature vector is obtained from the extracted patch. In the second layer, we transform the $n_1$-dimensional vector into $n_2$-dimensional one. This operation is equivalent to a filtration on the feature map from the first layer. The second can be formulated as

$$C_2(\mathbf{y}) = \text{ReLU}(\mathbf{W}_2 * C_1(\mathbf{y}) + \mathbf{b}_2), \qquad (4)$$

where $\mathbf{W}_2$ is composed of $n_2$ convolution kernels with a fixed size of $s_2 \times s_2$. If the desired network only has two layers, the output of this layer is the corresponding cleaned patches for the final reconstruction. Generally, inserting more layers is a way to potentially boost the capacity of the network. However, a deeper CNN is at a higher cost of computation including longer training time.

*2.2.3 Reconstruction*
In this step, the processed overlapping patches are merged into the final image. These overlapping patches are properly weighted before their summation. This operation can be

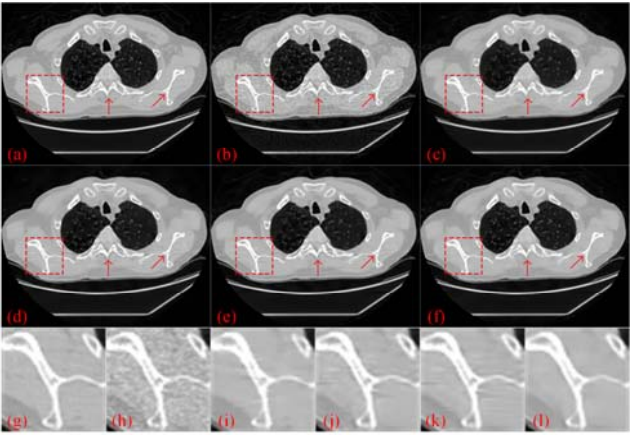

Fig. 2. Results of a slice of chest images. (a) Original normal-dose image; (b) the low-dose images; (c) the ASD-POCS reconstructed image; (d) the KSVD processed low-dose image; (e) the BM3D processed low-dose image; (f) the CNN processed low-dose image; and (g)-(l) the zoomed regions specified with the red boxes in (a)-(f).

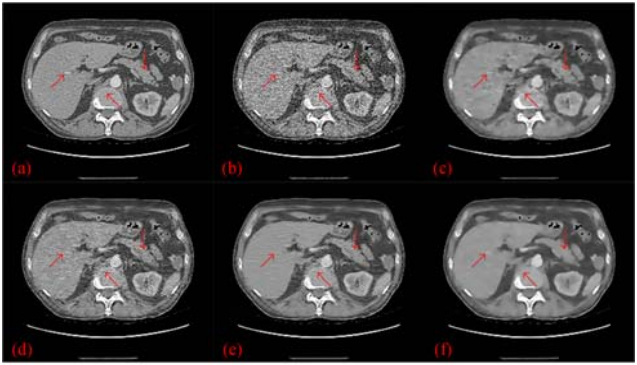

Fig. 3. Results of a slice of abdomen images. (a) Original normal-dose image; (b) the low-dose images; (c) the ASD-POCS reconstructed image; (d) the KSVD processed low-dose image; (e) the BM3D processed low-dose image; and (f) the CNN processed low-dose image.

considered as filtration by a pre-defined convolutional kernel formulated by

$$C(\mathbf{y}) = \mathbf{W}_3 * C_2(\mathbf{y}) + \mathbf{b}_3, \qquad (5)$$

where $\mathbf{W}_3$ is composed of only 1 convolution kernel with a size of $s_3 \times s_3$, and $\mathbf{b}_3$ has the same size as that of $\mathbf{W}_3 * C_2(\mathbf{y})$.

Eqs. (3)-(5) all use convolutional operations, although they have been designed for different purposes. That is the reason why the CNN architecture is in use for our low-dose CT image denoising.

### 2.2.4 Training

Once the network is configured, the parameter set, $\Theta = \{\mathbf{W}_1, \mathbf{W}_2, \mathbf{W}_3, \mathbf{b}_1, \mathbf{b}_2, \mathbf{b}_3\}$, of the network must be estimated to learn the function $C$. Given the training dataset $D = \{(\mathbf{x}_1, \mathbf{y}_1), (\mathbf{x}_2, \mathbf{y}_2), \ldots, (\mathbf{x}_N, \mathbf{y}_N)\}$ with $\{\mathbf{x}_i\}$ and $\{\mathbf{y}_i\}$ denoting normal-dose image patches and its corresponding noisy versions respectively, and $N$ being the total number of training samples, the estimation of the parameters can be achieved by minimizing the following loss function:

$$L(D; \Theta) = \frac{1}{N} \sum_{i=1}^{N} \|\mathbf{x}_i - C(\mathbf{y}_i)\|^2. \qquad (6)$$

The loss function is optimized using the stochastic gradient descent method.

## 3. EXPERIMENTS

To evaluate the performance of the proposed approach, the state-of-the-art methods, both iterative reconstruction and post-reconstruction processing, were selected for comparison, including ASD-POCS, KSVD and BM3D. The parameters for the completing methods were set according to the recommendation in the original references. The peak signal to noise ratio (PSNR), root mean square error (RMSE) and structure similarity index (SSIM) were used as the quantitative metrics. All the experiments were conducted using MATLAB 2015b on a PC (Intel i7 6700K CPU, 16 GB RAM and GTX 980 Ti graphics card).

Table I. Quantitative measurements associated with different algorithms.

| | | TV | KSVD | BM3D | CNN |
|---|---|---|---|---|---|
| Fig. 2 | PSNR | 41.22 | 40.6901 | 40.9998 | **41.6843** |
| | RMSE | 0.0087 | 0.0092 | 0.0089 | **0.0082** |
| | SSIM | 0.9527 | 0.9592 | 0.9657 | **0.9721** |
| Fig. 3 | PSNR | 37.5485 | 38.3841 | 38.9903 | **38.9904** |
| | RMSE | 0.0133 | 0.012 | 0.0112 | **0.0101** |
| | SSIM | 0.8825 | 0.9226 | **0.9295** | 0.9277 |
| All | PSNR | 41.5021 | 40.8445 | 41.5358 | **42.1514** |
| | RMSE | 0.0087 | 0.0096 | 0.0088 | **0.0080** |
| | SSIM | 0.9447 | 0.9509 | 0.9610 | **0.9707** |

### 3.1 Dataset preparation

Totally 7,015 CT normal-dose images with size of 256×256 from 165 patients including different parts of human body were downloaded from NCIA (National Cancer Imaging Archive). Fig. 1 illustrates typical slices included in our training set.

The corresponding low-dose images were generated by imposing Poisson noise into each detector element of the simulated normal-dose sinogram with the blank scan flux $b_0 = 10^5$. Siddon's ray-driven algorithm was used in fan-beam geometry. The source-to-rotation center distance was 40 cm while the detector-to-rotation center was 40 cm. The image size was 20 cm × 20 cm. The detector width was 41.3 cm in length containing 512 detector elements. The data were uniformly sampled in 1024 views over a full scan. The input patches of the network were extracted from the original images with size $m = 33$. The slide step was 4. The original 200 training images included in our first experiment resulted in about $10^6$ samples. There are two reasons why patches were used, instead of whole images: one is that the images can be well represented by local structures; and the other is that deep learning requires a big training dataset and chopping the original images into patches can efficiently boost the number of samples.

For fairness, in the testing stage, 100 low-dose images were randomly selected from the original dataset to form the testing set, and 200 images excluding the ones from the same patients involved in testing set were randomly selected as the training set.

## 3.2 Parameter setting

In this paper, three layers were used in the proposed network. The filter number, $n_1$ and $n_2$, were respectively set to 64 and 32, and the corresponding filter sizes, $s_1$, $s_2$ and $s_3$, were set to 9, 3 and 5 respectively. The initial weights of the filters in each layer were randomly set, which satisfies the Gaussian distribution with zero mean and standard deviation 0.001. The initial learning rate was 0.001 and slowly decayed to 0.0001 during the training process.

## 3.3 Results

### 3.3.1 Visual inspection

For this purpose, we selected 2 representative slices through the chest and abdomen respectively. Figs. 2 and 3 gives the results from different methods. In both the figures, the noise and artifacts caused by the lack of incident photons severely degraded the image quality. Some details and structures cannot be discriminated. All of the methods can eliminate the noise and artifacts to different degrees. However, due to the piecewise constant assumption by the TV minimization, ASD-POCS caused blocky effects in the resultant images. KSVD and BM3D could not efficiently suppress the streak artifacts near the bone as indicated by the red arrow in Fig. 2 and the zoomed parts (Fig. 2 (g)-(l)) can show more details. In Fig. 3, the arrows point to small details and boundaries where only the proposed CNN method could recover the most details.

### 3.3.2 Quantitative measurement

To quantitatively evaluate the proposed CNN algorithm, PSNR, RMSE and SSIM were measured for all the images in the testing set. The results for the restored images in Figs. 2 and 3 are in Table I. It can be seen in Table I that for Fig. 2 our proposed method obtained the best results, which are consistent to the visual inspection. In Fig. 3, CNN also achieved the best performance except for SSIM. One possible reason for this small discrepancy in terms of SSIM is that most part of the abdomen image is soft tissue and BM3D is more preferable in this situation. The term "All" in Table I means the average values of the measurements for all the 100 images in the testing set. It can be observed that all the metrics demonstrate the proposed CNN based method had the best performance.

## 4. CONCLUSION

In this pilot study, we have evaluated the performance of a deep convolutional neural network for noise reduction in low-dose CT. The results demonstrate the potential of CNN based method for medical imaging. In the future, the proposed network structure will be optimized and applied to other CT topics, such as few-view CT reconstruction, metal artifact reduction, and interior CT. Another possible direction is to investigate other network architectures to deal with dynamic and spectral CT problems.